# Unconventional Fermi surface in an insulating state

B. S. Tan,[1] Y. -T. Hsu,[1] B. Zeng,[2] M. Ciomaga Hatnean,[3] N. Harrison,[4] Z. Zhu,[4]
M. Hartstein,[1] M. Kiourlappou,[1] A. Srivastava,[1] M. D. Johannes,[5] T. P. Murphy,[2]
J. -H. Park,[2] L. Balicas,[2] G. G. Lonzarich,[1] G. Balakrishnan,[3] Suchitra E. Sebastian[1*]

[1]Cavendish Laboratory, Cambridge University, JJ Thomson Avenue, Cambridge CB3 OHE, U.K.,
[2]National High Magnetic Field Laboratory, Tallahassee, FL 32310,
[3]Department of Physics, University of Warwick, Coventry, CV4 7AL, U.K.,
[4]National High Magnetic Field Laboratory, LANL, Los Alamos, NM 87504,
[5]Center for Computational Materials Science, Naval Research Laboratory, Washington, DC 20375

**Insulators occur in more than one guise, a recent finding was a class of topological insulators, which host a conducting surface juxtaposed with an insulating bulk. Here we report the observation of an unusual insulating state with an electrically insulating bulk that simultaneously yields bulk quantum oscillations with characteristics of an unconventional Fermi liquid. We present quantum oscillation measurements of magnetic torque in high purity single crystals of the Kondo insulator $SmB_6$, which reveal quantum oscillation frequencies characteristic of a large three-dimensional conduction electron Fermi surface similar to the metallic rare earth hexaborides such as $PrB_6$ and $LaB_6$. The quantum oscillation amplitude strongly increases at low temperatures, appearing strikingly at variance with conventional metallic behaviour.**

Kondo insulators, a class of materials positioned close to the border between insulating and metallic behaviour, provide fertile ground for unusual physics [1, 2, 3, 4, 5, 6, 7, 8, 9, 10, 11, 12, 13, 14]. This class of strongly correlated materials is thought to be characterised by a ground state with a small energy gap at the Fermi energy owing to the collective hybridisation of conduction and $f$-electrons. The observation of quantum oscillations has traditionally been associated with a Fermi liquid state; here we present the surprising measurement of quantum oscillations in the Kondo insulator $SmB_6$ [15] that originate from a large three-dimensional Fermi surface occupying half the Brillouin zone and strongly resembling the conduction electron Fermi surface in the metallic rare earth hexaborides [16]. Our measurements in $SmB_6$ reveal a dramatic departure from conventional metallic Lifshitz Kosevich behaviour [18]; instead of the expected saturation at low temperatures, a striking increase is observed in the quantum oscillation amplitude at low temperatures .

Single crystals of $SmB_6$ used in the present study were grown by means of the image furnace technique [19] in order to achieve high purities as characterised by the high inverse residual resistivity ratio. Single crystals with inverse resistance ratios [IRR $= R(T = 1.8 \, \text{K})/R(T = 300 \, \text{K})$, where $R$ is resistance and $T$ is temperature] of the order of $10^5$ were selected for this study; the IRR has been shown to characterise crystal quality, with the introduction of point defects by radiation damage [20], or through off-stoichiometry [21] resulting in a decrease in low temperature resistance and an increase in high temperature resistance. The resistance of a $SmB_6$ single crystal is shown in Fig. 1B measured as a function of temperature at zero magnetic field and in an applied DC magnetic field of 45 T, demonstrating that activated electrical conductivity characteristic of an energy gap $\approx 40$ K at the Fermi energy persists up to high magnetic fields. The non-magnetic ground state of $SmB_6$ is evidenced by the linear magnetisation up to 60 T (Fig. 1B, bottom inset).

We observed quantum oscillations in $SmB_6$ by measuring the magnetic torque. The measurements were done in magnetic fields up to 40 T and down to $T = 0.4$ K, and in magnetic fields up to 35 T and down to $T = 0.03$ K. Quantum oscillations periodic in inverse magnetic field are observed against a quadratic background, with frequencies ranging from 50 to 15,000 T

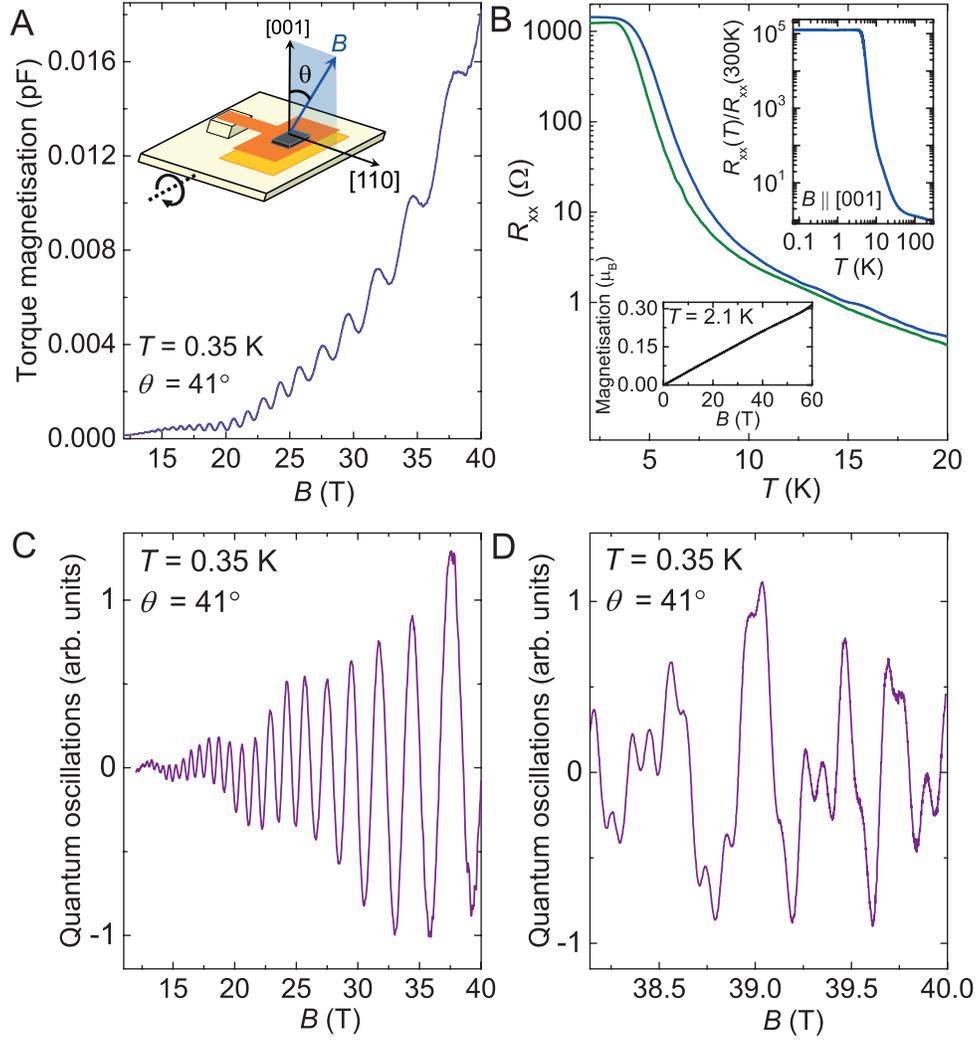

Fig. 1: **Quantum oscillations in the magnetic torque in SmB$_6$.** (A) Quantum oscillations in magnetic torque are visible against a quadratic background. (Inset) Schematic of the magnetic torque measurement setup using a capacitive cantilever and the notation for angular rotation by angle $\theta$. (B) Resistance as a function of temperature in zero magnetic field (blue line), and at 45 T (green line) using an unchanged measurement configuration on a SmB$_6$ sample of dimensions 1.1 by 0.3 by 0.1 mm. (Top inset) Measured resistance from 80 mK up to high temperatures, from which the high IRR can be ascertained [a fit to activated electrical conductivity is provided in [23]]. (Bottom inset) Magnetisation of SmB$_6$ at 2.1 K remains linear up to 60 T. (C) Dominant low frequency quantum oscillations can be discerned after background subtraction of a sixth order polynomial. (D) Magnetic torque at the highest measured fields after the subtraction of the low frequency background torque. Quantum oscillations are visible in an intermediate frequency range (between 2,000 and 4,000 T) as well as a high frequency range up to 15,000 T.

(Fig. 1A,C, and D). A Fourier transform of the quantum oscillations is shown in Figure 2A as a function of inverse magnetic field, revealing well-defined peaks corresponding to multiple frequencies. The periodicity of the quantum oscillations in inverse magnetic field is revealed by the linear Landau index plot in Fig. 2B.

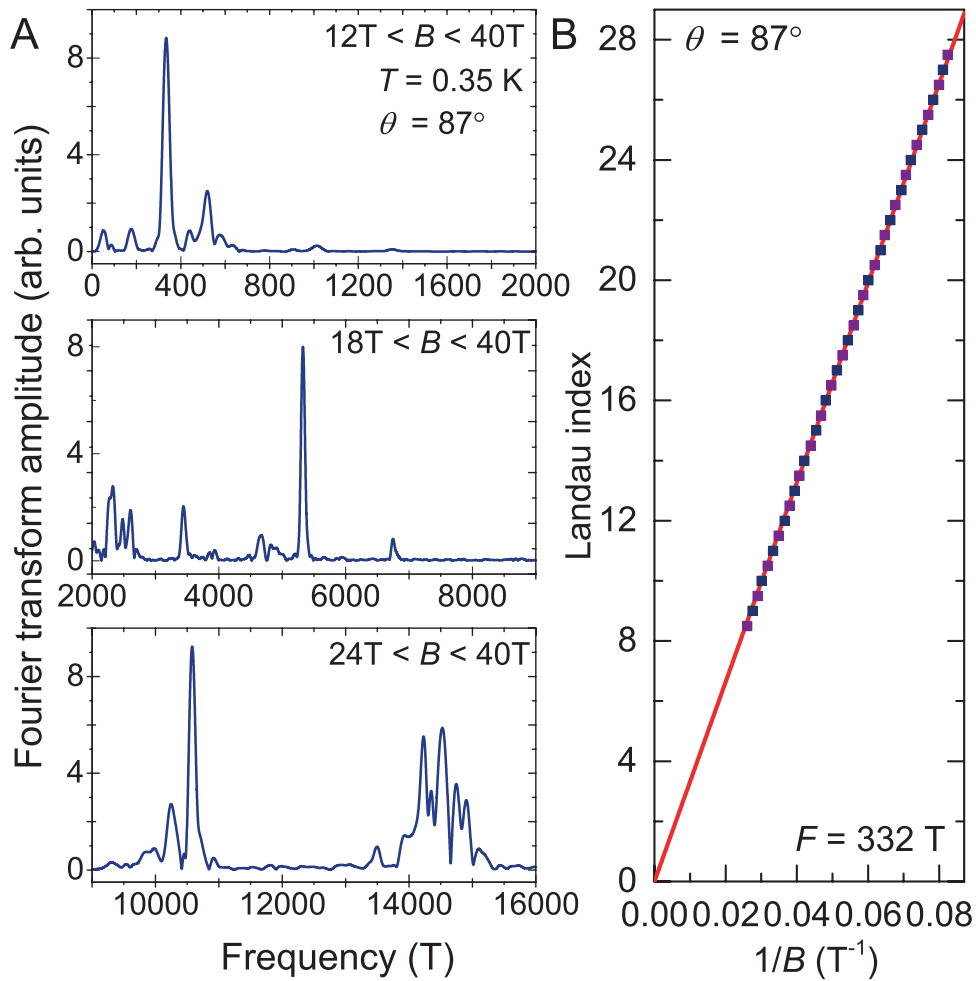

Fig. 2: **Landau quantisation in SmB$_6$.** (A) Fourier transforms of magnetic torque as a function of inverse magnetic field, from which a polynomial background has been subtracted, revealing multiple quantum oscillatory frequencies ranging from 50 T to 15,000 T. Field ranges for analysis have been chosen that best capture the observed oscillations, with the highest frequencies only appearing in the higher field ranges. (B) The maxima and minima in the derivative of magnetic torque with respect to the magnetic field, corresponding to the dominant low frequency oscillation, are plotted as a function of inverse magnetic field; the linear dependence signals Landau quantisation.

The observation, especially of rapid quantum oscillations with frequencies higher than 10 kT (corresponding to approximately half the volume of the cubic Brillouin zone) in $SmB_6$, is striking. This observation is in contrast to previous reports of very low frequency quantum oscillations corresponding to a few percent of the Brillouin zone in $SmB_6$, attributed to a two-dimensional surface contribution [22]. Our observation of very high quantum oscillation frequencies requiring mean free paths on the order of a few micrometers would be challenging to explain from a surface layer of a few atomic lengths thickness, which would typically yield such rapid frequencies only at a special angle of inclination at which the cyclotron orbit lies completely within the surface layer. Key to identifying the Fermi surface from which the observed quantum oscillation frequencies originate is a comparison with previous quantum oscillation measurements on metallic hexaborides such as nonmagnetic $LaB_6$, antiferromagnetic $CeB_6$, and antiferromagnetic $PrB_6$ [16, 17]. These materials exhibit a metallic ground state involving predominantly conduction electrons, with a low residual resistivity of the order of one microhm.cm ($\approx 10^6$ times lower than in Kondo insulating $SmB_6$), and are characterised by a multiply connected Fermi surface of prolate spheroids (Fig. 3, D and E). Strikingly, the angular dependence of the various quantum oscillation frequencies in $SmB_6$ reveals characteristic signatures of the three-dimensional Fermi surface identified in the metallic rare-earth hexaborides (Fig. 3, A to C). In particular, the high observed $\alpha$ frequencies (Fig. 3A) reveal the characteristic symmetry of large prolate spheroids centred at X-points of the Brillouin zone (Fig. 3,D and E), whereas the lower observed frequencies (Fig. 3A) reveal the characteristic symmetry of small ellipsoids located at the neck positions. Both of these types of ellipsoids are universal Fermi surface features identified from experiment and band structure calculations in the metallic rare-earth hexaborides [16, 17]. Similar features are also revealed in density functional calculations of $SmB_6$ when the Fermi energy is shifted from its calculated position in the insulating gap either up into the conduction or down into the valence bands (Fig. 3,D and E).

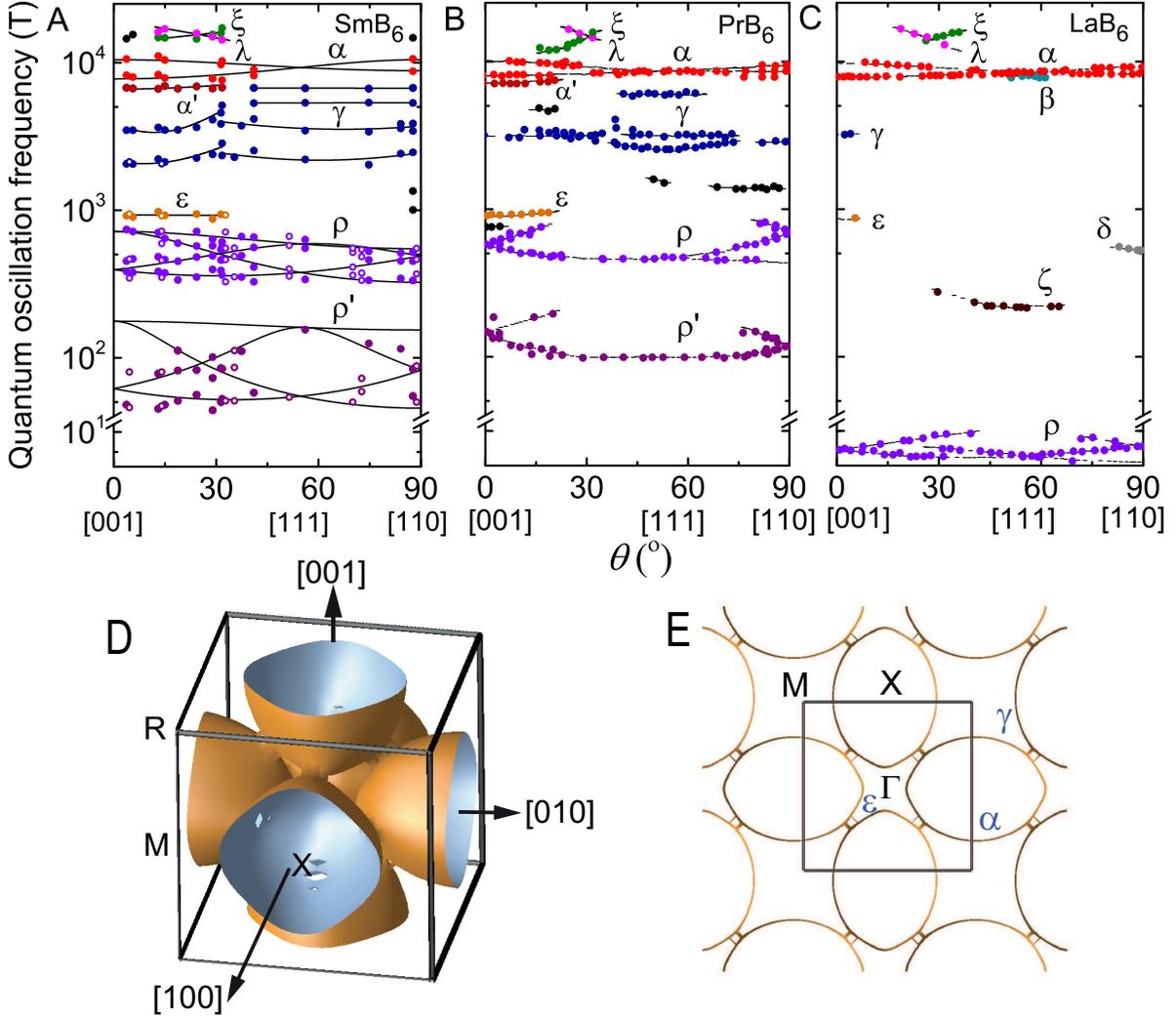

Fig. 3: **Angular dependence of the quantum oscillation frequencies in SmB$_6$.** (A) Data from two of the SmB$_6$ samples in which oscillations were observed are shown, indicated by solid and open circles. One of the samples (solid circles) was prepared as a thin plate with the dominant face perpendicular to the [100] axis [sample 1 [23]]. (B and C) The second sample (open circles) was prepared as a thin plate with the dominant face perpendicular to the [110] axis (sample 2, [23]). The angular dependence strongly resembles that of the three-dimensional Fermi surface in antiferromagnetic PrB$_6$ shown in (B), and nonmagnetic LaB$_6$ shown in (C) [17]. (D and E) The $\alpha$ orbit in red in all the rare earth hexaborides is fit to large multiply connected prolate spheroids centred at the X points of the Brillouin zone, shown in (D); a cross-section in the XM plane is shown in (E). The $\rho$ and $\rho'$ orbits in each of the rare earth hexaborides are fit to small ellipsoids located at the neck positions [not shown in (D) and (E)]. More details of the fits are provided in [23]. The remaining intermediate orbits are shown with lines as a guide to the eye. All orbit identifications have been made after measured frequencies and band structure calculations in PrB$_6$ and LaB$_6$ [17]. (D) and (E) show Fermi surfaces calculated for SmB$_6$ using density functional theory [23], with a downward shift of the Fermi energy from its calculated position within the gap to expose the unhybridised bands, and yield pocket sizes similar to experiment.

The observed angular dependence of quantum oscillations in $SmB_6$ remains the same irrespective of whether the sample is prepared as a thin plate with a large plane face perpendicular to the [110] direction, or to the [100] direction (fig. S1), and exhibits the same characteristic signatures with respect to the orientation of the magnetic field to the crystallographic symmetry axes of the bulk crystal (Fig. 3A). The bulk quantum oscillations we measure in $SmB_6$ corresponding to the three-dimensional Fermi surface mapped out in the metallic rare-earth hexaborides may not be directly related to the potential topological character of $SmB_6$, which would have as its signature a conducting surface [24]. In addition to the magnetic torque signal from the atomically thin surface region being several orders of magnitude smaller than the signal from the bulk, the observation of surface quantum oscillations would be rendered more challenging by the reported Sm depletion and resulting reconstruction of Sm ions at the surface layer of $SmB_6$ [25].

The unconventional character of the state we measure in $SmB_6$ becomes apparent upon investigating the temperature dependence of the quantum oscillation amplitude in $SmB_6$. We found that between $T = 25$ K and 2 K, the quantum oscillation amplitude exhibits a Lifshitz-Kosevich like temperature dependence (Fig. 4), characteristic of a low effective mass similar to that of metallic $LaB_6$, which has only conduction electrons [16]. The comparable size of low temperature electronic heat capacity measured for our $SmB_6$ single crystals to that of metallic $LaB_6$ [23] also seems to suggest a large Fermi surface with low effective mass in $SmB_6$. However, instead of saturating at lower temperatures as would be expected for the Lifshitz Kosevich distribution characteristic of quasiparticles with Fermi Dirac statistics [18], the quantum oscillation amplitude increases dramatically as low temperatures down to 30 mK are approached (Fig. 4). Such non-Lifshitz Kosevich temperature dependence is remarkable, given the robust adherence to Lifshitz Kosevich temperature dependence in most examples of strongly correlated electron systems, from the underdoped cuprate superconductors [26], to heavy fermion systems [27, 28], to systems displaying signatures of quantum criticality [29], a notable exception being fractional quantum Hall systems [30, 31]. The possibility of a subtle departure from Lifshitz Kosevich temperature dependence has been reported in a few materials [32, 33].

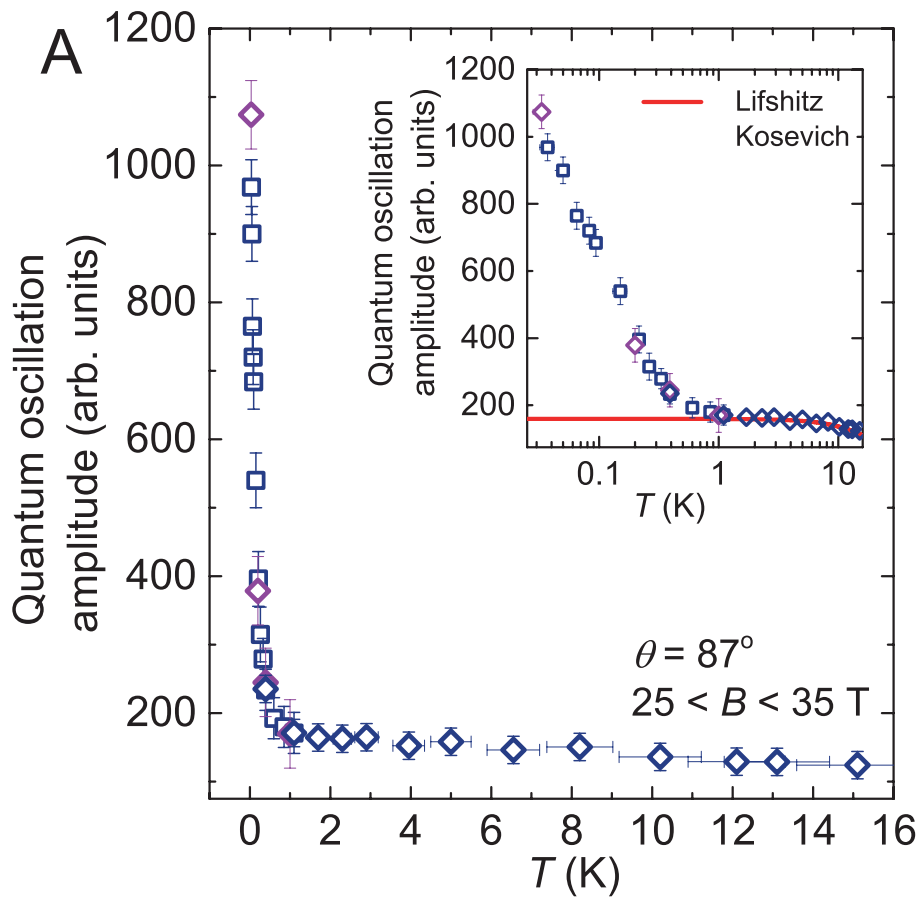

Fig. 4: **Temperature dependence of quantum oscillation amplitude.** The dominant 330 T frequency over the magnetic field range 25 to 35 T is shown, revealing a steep increase in amplitude at low temperatures. The measurements in the temperature range from 25 K down to 0.35 K were performed in a $^3$He fridge in the hybrid magnet [sample 1 [23], blue diamonds], whereas the measurements at temperatures in the range from 1 K down to 30 mK were done in a dilution fridge in the resistive magnet on two different samples [sample 1, purple diamonds; sample 3 [23]: blue squares]. At low temperatures, a strong deviation from the conventional Lifshitz Kosevich form can be seen in the inset by comparison with a simulated Lifshitz Kosevich form for effective mass $m^* = 0.18\, m_\mathrm{e}$. A logarithmic temperature scale is used in the inset for clarity.

The ground state of SmB$_6$ is fairly insensitive to applied magnetic fields, with activated electrical conductivity behaviour across a gap remaining largely unchanged up to at least 45 T (Fig. 1B). Such a weak coupling to the magnetic field is in contrast to unconventional states in other materials that are tuned by an applied magnetic field [6, 7, 8, 9, 10, 11, 13]. Furthermore, this rules out the possibility of quantum oscillations in $SmB_6$ arising from a high magnetic field state in which the energy gap is closed. The possibility that quantum oscillations arise from static, spatially disconnected metallic patches of at least 1 $\mu$m length scale that do not contribute to the electrical transport also appears unlikely. Similar quantum oscillations are observed in all (more than 10) measured high quality samples in multiple high magnetic field experiments, with the best samples yielding magnetic quantum oscillations of amplitude corresponding to a substantial fraction of the expected size from bulk $SmB_6$. The presence of rare-earths other than Sm has been ruled out to within $0.01$ % by means of chemical analysis and scanning electron microscopy [23]. Off-stoichiometric metallic regions of $SmB_6$ appear an unlikely explanation for our results, given reports that up to 30% Sm depletion does not close the energy gap [20], whereas scanning electron microscopy of our samples reveals a homogeneity of within 1 % of Sm concentration over the sample area [23]. The possibility of spatially disconnected strained regions of $SmB_6$, which is known to become metallic under applied pressures on the order of 10 GPa, or static spatially disconnected islands of hybridised and unhybridised Sm $f$-electrons also seems unlikely. An improvement in the IRR by means of the removal of strain with electropolishing strengthens the quantum oscillation signal whereas straining the sample by means of thermal cycling weakens the quantum oscillation signal [23]. Further, the interplay between hybridised and unhybridised Sm $f$-electrons which may be an important ingredient in the physics of $SmB_6$, has been revealed by Mössbauer and muon-spin relaxation experiments to be homogenous and dynamically fluctuating, rather than being manifested as static spatially inhomogeneous regions [34, 35].

The insulating state in $SmB_6$ in which low energy excitations lack long range charge trans-port as shown by the activated dc electrical conductivity, but display extended character as shown by quantum oscillations, poses a mystery. A clue might be provided by slow fluctuations between a collectively hybridized insulating state and an unhybridized state in which the conduction electrons form a solely conduction electron Fermi surface, similar to that we observe [2, 35, 36, 37, 38, 39]. A fluctuation timescale in the range between $10^{-8}$ and $10^{-11}$ seconds is suggested by previous x-ray absorption spectroscopy and Mössbauer measurements [40]. This timescale is longer or comparable with the inverse cyclotron frequency ($1/\omega_c$) which is on the order of $10^{-11}$ seconds for the measured cyclotron orbits. Intriguingly, similar slow fluctuations have been invoked to explain quantum critical signatures in the metallic $f$-electron system $\beta$-YbAlB$_4$ [41]. SmB$_6$ may be viewed as being on the border of quantum criticality in the sense that it transforms from a non-magnetic insulating phase to a magnetic metallic phase under applied pressures on the order of 10 GPa [42, 43, 44, 45], which is in contrast to other metallic rare earth hexaborides in which the $f$-electrons order magnetically in the ambient ground state. Our observation of a large three-dimensional conduction electron Fermi surface revealed by quantum oscillations may be related to reports of a residual density of states at the Fermi energy in SmB$_6$ through measurements of heat capacity [23, 46], optical conductivity [47], Raman scattering [48], and neutron scattering [49]. Another possibility is that quantum oscillations could arise even in a system with a gap in the excitation spectrum at the Fermi energy, provided that the size of the gap is not much larger than the cyclotron energy [50]. Within this scenario, the residual density of states observed at the Fermi energy with complementary measurements, and the steep upturn in quantum oscillation amplitude we observe at low temperatures appear challenging to explain.

# Acknowledgements

B.S.T., Y.T.H., M.H., M.K., A.S., and S.E.S. acknowledge support from the Royal Society, the Winton Programme for the Physics of Sustainability, and the European Research Council (ERC) under the European Unions Seventh Framework Programme (grant FP/2007-2013)/ERC Grant Agreement 337425. B.Z. and L.B. acknowledge support from the U.S. Department of Energy (DOE) - Basic Energy Sciences (BES) through award DE-SC0002613. M.C.H. and G.B. acknowledge support from Enginerring and Physical Sciences Research Council (EPSRC) grant EP/L014963/1. N.H. and Z.Z. acknowledge support from the DOE Office of Science, BES - Materials Science and Engineering "Science of 100 Tesla" programme. MDJ acknowledges support for this project by the Office of Naval Research (ONR) through the Naval Research Laboratory's Basic Research Program. GGL acknowledges support from EPSRC grant EP/K012894/1. A portion of this work was performed at the National High Magnetic Field Laboratory, which is supported by NSF Cooperative Agreement DMR-1157490 and the state of Florida. We acknowledge valuable inputs from G. Baskaran, D. Benkert, A. K. Cheetham, D. Chowdhury, P. Coleman, N. R. Cooper, M. P. M. Dean, O. Ertem, J. Flouquet, R. H. Friend, R. Golombok, C. Harris, S. A. Hartnoll, T. Kasuya, G. Khaliullin, E. -A. Kim, J. Knolle, P. A. Lee, P. B. Littlewood, C. Liu, K. Miyake, J. E. Moore, O. Petrenko, S. Sachdev, A. Shekhter, N. Shitsevalova, Q. Si, A. Thomson, S. Todadri, C. M. Varma, and J. Zaanen. We thank magnet laboratory personnel including J. Billings, R. Carrier, E. S. Choi, B. L. Dalton, D. Freeman, L. J. Gordon, M. Hicks, C. H. Mielke, J. M. Petty, and J. N. Piotrowski for their assistance.